\documentstyle[twoside,fleqn,epsfig]{article}

\def\be{\begin{equation}}
\def\ee{\end{equation}}
\def\bea{\begin{eqnarray}}
\def\eea{\end{eqnarray}}
\newcommand{\lsim}{\mathrel{\mathop{\kern 0pt \rlap
  {\raise.2ex\hbox{$<$}}} \lower.9ex\hbox{\kern-.190em $\sim$}}}
\newcommand{\gsim}{\mathrel{\mathop{\kern 0pt \rlap
  {\raise.2ex\hbox{$>$}}}
  \lower.9ex\hbox{\kern-.190em $\sim$}}}

\newcommand{\AmS}{{\protect\the\textfont2
  A\kern-.1667em\lower.5ex\hbox{M}\kern-.125emS}}
\normalsize
\begin{document}

\baselineskip=0.65cm

\begin{center}
\Large{\bf Investigating Earth shadowing effect with DAMA/LIBRA--phase1}
\rm
\end{center}

\large

\begin{center}

\vskip 5mm

R.\,Bernabei,~P.\,Belli,~S.\,d'Angelo\footnote{Deceased}, A. Di Marco,~F.\,Montecchia$^{a}$
\vspace{2mm}

{\small {\it Dip. di Fisica, Universit\`a di Roma ``Tor Vergata'', I-00133  Rome, Italy}} \\
{\small {\it I.N.F.N., sez. Roma ``Tor Vergata'', I-00133 Rome, Italy}} \\
{\small {\it $^{a}$ also in: Dip. di Ingegneria Civile e Ingegneria Informatica, Universit\`a di Roma
``Tor Vergata'', I-00133  Rome, Italy}} \\
\vspace{4mm}

A.\,d'Angelo,~A.\,Incicchitti
\vspace{2mm}

{\small {\it Dip. di Fisica, Universit\`a di Roma ``La Sapienza'', I-00185 Rome, Italy}} \\
{\small {\it I.N.F.N., sez. Roma, I-00185 Rome, Italy}} \\
\vspace{4mm}

F.\,Cappella,~V.\,Caracciolo,~R.\,Cerulli
\vspace{2mm}

{\small {\it Laboratori Nazionali del Gran Sasso, I.N.F.N., Assergi, Italy}} \\
\vspace{4mm}

C.J.\,Dai,~H.L.\,He,~H.H.\,Kuang,~X.H.\,Ma, X.D.\,Sheng,~R.G.\,Wang and ~Z.P.\,Ye$^{b}$
\vspace{2mm}

{\small {\it Key Laboratory of Particle Astrophysics, Institute of High Energy Physics, Chinese
Academy of Sciences, P.O. Box 918/3, 100049 Beijing, China}} \\
{\small {\it $^{b}$ also in: University of Jing Gangshan, Jiangxi, China}}

\end{center}
\vskip 5mm

\begin{abstract}

In the present paper the results obtained in the investigation of possible diurnal effects for low-energy {\it single-hit} scintillation 
events of DAMA/LIBRA--phase1 (1.04 ton $\times$ yr exposure) have been analysed in terms of an effect 
expected in case of Dark Matter (DM) candidates inducing nuclear recoils and having high 
cross-section with ordinary matter, which implies low DM local 
density in order to fulfill the DAMA/LIBRA DM annual modulation results. 
This effect is due to the different Earth depths crossed by those DM candidates  
during the sidereal day.
\end{abstract}

\vspace{5.0mm}


\vspace{2.0mm}


\section{Introduction}

The present DAMA/LIBRA experiment 
\cite{perflibra,modlibra,modlibra2,modlibra3,bot11,pmts,mu,review,papep,cnc-l,IPP,diurna,norole}, as the former DAMA/NaI 
\cite{prop,que,allDM2,allDM3,allDM4,allDM5,allDM6,allDM7,halo,Nim98,Sist,RNC,ijmd,ijma,epj06,ijma07,chan,wimpele,ldm,allRare1,allRare2,allRare3,allRare4,allRare5,allRare6,allRare7,allRare8,allRare9,allRare10,IDM96} 
has the main aim to investigate the presence of DM particles in the galactic halo by exploiting the 
model-independent DM annual modulation signature (originally suggested in Refs. \cite{Drukier,Freese}).
In particular, they have cumulatively reached a model independent evidence at 9.3$\sigma$ C.L. for the 
presence of DM particles in the galactic halo by exploiting the DM annual modulation 
signature \cite{modlibra3}. Recently the results obtained by investigating the presence 
of possible diurnal variation in the low-energy {\it single-hit} scintillation events collected by
DAMA/LIBRA--phase1 (1.04 ton $\times$ yr exposure) have been released and analysed in terms of a DM
second order model-independent effect due to the Earth diurnal rotation around its axis \cite{diurna}. 
In particular, the data were analysed using the sidereal time referred to Greenwich, often called 
GMST. No diurnal variation with sidereal time has been observed at the reached level of sensitivity, 
which was not yet adequate to point out the effect searched for there.
In the present paper those experimental data are analysed in terms of an effect -- named ``{\it Earth Shadow Effect}'' --
 which could be expected for DM candidate particles inducing nuclear recoils; this effect would be induced by the variation 
-- during the day -- of the Earth thickness crossed by the DM particle in order to reach the experimental 
set-up. It is worth noting that a similar effect can be pointed out only for candidates with high cross-section with ordinary matter, which implies 
low DM local density in order to fulfill the DAMA/LIBRA DM annual modulation results.  
Such DM candidates could get trapped in substantial quantities in the Earth's core; in this case
they could annihilate and produce secondary particles (e.g. neutrinos) and/or they could carry thermal energy away from the core, 
giving potentiality to further investigate them.

Preliminary investigations on DM candidates inducing diurnal variation
were performed in Refs. \cite{Col92,Col93,Nim98} and more recently in
Ref. \cite{altr_diu}. 

\section{The {\it Earth Shadow Effect}}

During a sidereal day the Earth shields a terrestrial 
detector with a varying thickness, and this induces a variation of the flux of the DM candidates impinging the detector, mainly because of 
the modification of their velocity distribution, $f(\vec{v})$. It is worth noting that this {\it Earth Shadow Effect} is very 
small and could be detectable only in case of candidates with high cross-section with ordinary matter (i.e. present
in the galactic halo with small abundance).

The detector (and the hosting laboratory) velocity in the Galactic frame can be written as:
\begin{equation}
\vec{v}_{lab} (t) = \vec{v}_{LSR} + \vec{v}_{\odot} + \vec{v}_{rev}(t) + \vec{v}_{rot}(t), 
\label{eq:vlab}
\end{equation}
where: (i) $\vec{v}_{LSR}$ is the velocity of the Local Standard of Rest (LSR) because of the rotation of the 
Galaxy; (ii) $\vec{v}_{\odot}$ is the Sun peculiar velocity with respect to LSR; 
(iii) $\vec{v}_{rev}(t)$ is the velocity of the orbital motion of the Earth around the Sun and 
(iv) $\vec{v}_{rot}(t)$ is the velocity of the rotation of the Earth around its axis.
The two latter terms change as function of the sidereal time, $t$.
Using the galactic coordinate frame (that is $x$ axis towards the galactic center,
$y$ axis following the rotation of the Galaxy and the $z$ axis towards the galactic North pole), one gets:
$\vec{v}_{LSR}=(0,v_0,0)$, where $v_0= (220\pm50)$ km/s (uncertainty at 90\% C.L.) \cite{halo,astr_v0} is the local 
velocity, and
$\vec{v}_{\odot} = (9, 12, 7)$ km/s \cite{delh65}.

The DM particles in the galactic halo have a velocity distribution $g(\vec{w})$, which depends on the 
considered galactic halo model. Ref. \cite{halo} has shown many possible scenarios for 
the galactic halo; in the following we consider the isothermal halo model just because of its simplicity:
\begin{equation}
g(\vec{w}) = A e^{-\frac{w^2}{v_0^2}} \theta (v_{esc} - |\vec{w}|), 
\label{eq:v}
\end{equation}
with $A$ normalization constant and ${v}_{esc}$ escape velocity, assumed in the following equal to
650 km/s, as often considered in literature; however, it is also affected by uncertainty. 
However, no sizeable differences are observed in the outcome when a different value of $v_{esc}=550\, \rm km/s$ 
is considered, more closer to the 90\% C.L. range of the RAVE Survey results \cite{rave07}.
In the laboratory frame the DM velocity distribution $f(\vec{v})$ is obtained from eq. \ref{eq:v} 
straightforward since $\vec{w}=\vec{v}+\vec{v}_{lab}$.

\begin{figure}[!ht]
\begin{center}
\includegraphics[width=0.35\textwidth]{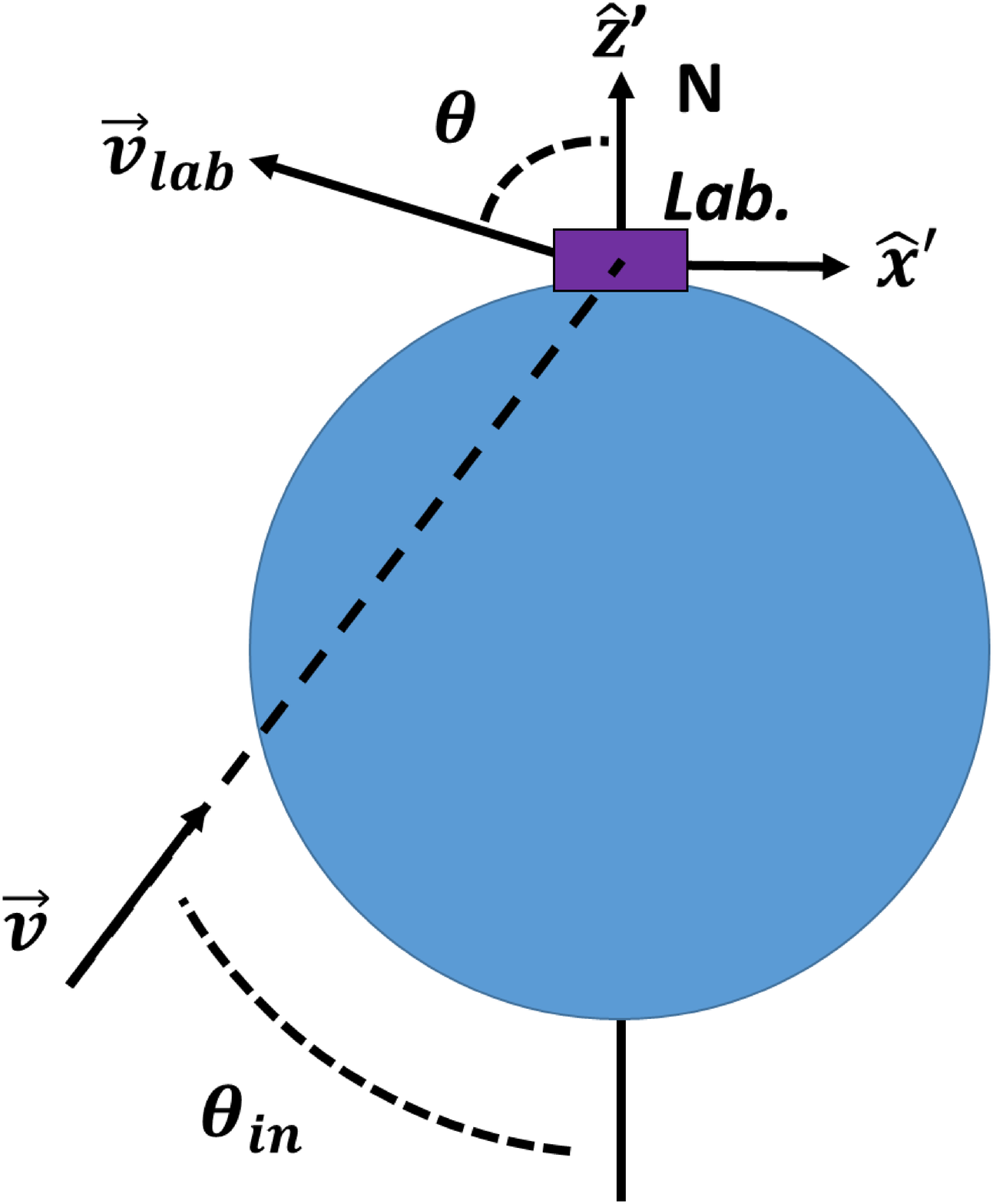}
\includegraphics[width=0.55\textwidth]{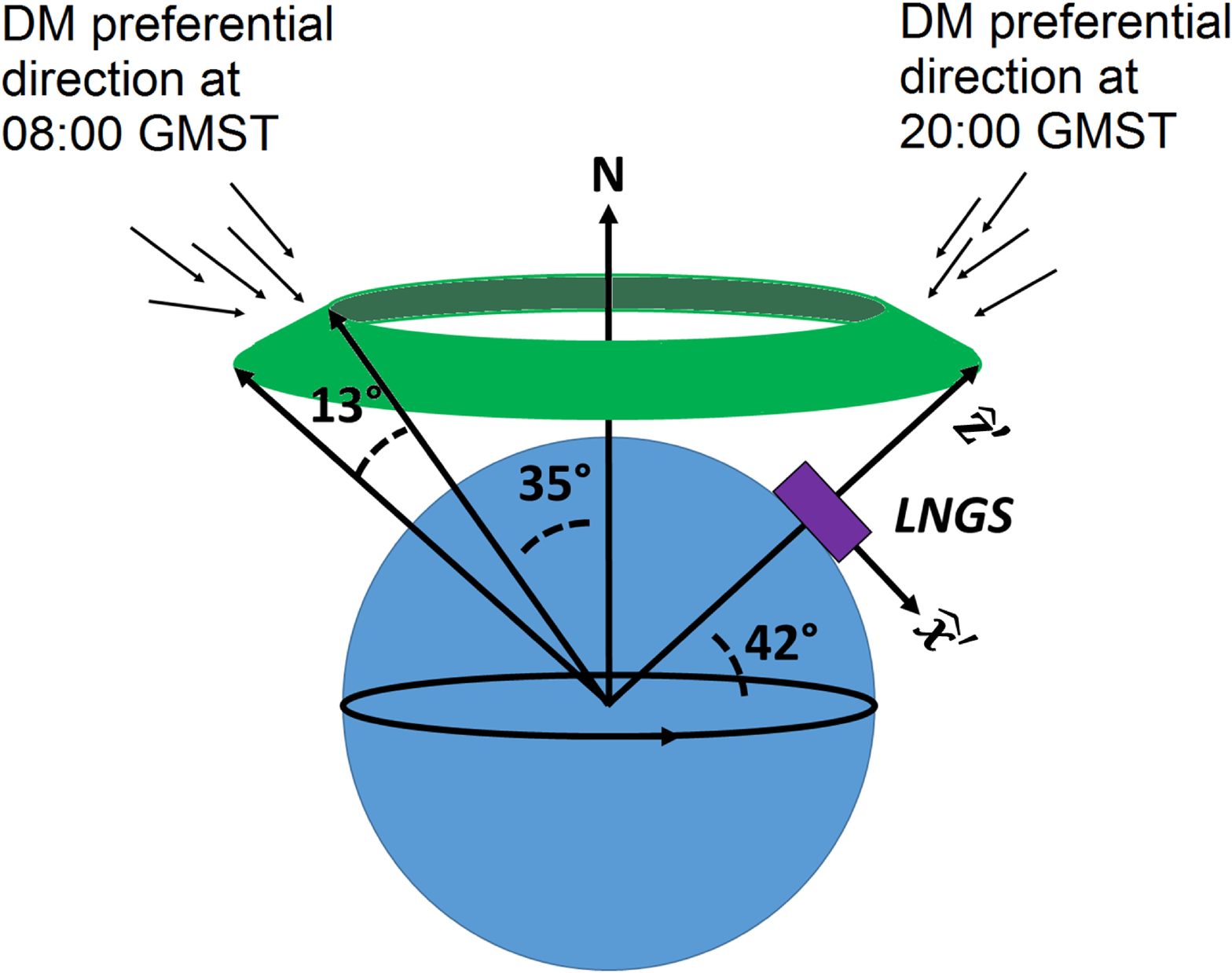}
\end{center}
\caption{Schematic view of the DM particles impinging direction on a detector;
the $x', y', z'$ represent the laboratory frame coordinate system. {\it Left:} schematic 
representation of the correlation between the thickness, $d$, crossed by the considered DM candidates to 
reach a laboratory (hypotetically placed at the geographic North pole) 
and the DM impinging angle, $\theta_{in}$. {\it Right:} schematic representation 
of the experimental condition considered in the text: detector placed at the Gran 
Sasso National Laboratory (LNGS) with the $z'$ axis in the vertical direction and the $x'$ axis pointing to the 
vernal equinox.}
\label{fig:fig1}
\end{figure}

To evaluate the expected daily variation of the DM particles velocity distribution due to the
{\it Earth Shadow Effect}, it is necessary to estimate the time dependence of the $\theta$ angle, 
the ``zenith distance'' of $\vec{v}_{lab}$ (i.e. the distance between $\vec{v}_{lab}$ and the zenith, 
see Fig. \ref{fig:fig1}). This can be determined by astrophysical considerations studying the Earth's rotation 
around its axis.

The simplest way to calculate $\theta(t)$ is in the equatorial coordinate system \cite{diurna} where the 
$\hat{e}_1^{ecs}$ axis is directed towards the vernal equinox, and $\hat{e}_1^{ecs}$ and $\hat{e}_2^{ecs}$ are 
on the equatorial plane; the $\hat{e}_3^{ecs}$ axis is towards the North pole. The right-handed convention is 
used. To work out the galactic coordinates of those versors, one considers: i) the equatorial coordinates of 
the galactic North pole: $RA = 192^\circ.859508$ and $DE = 27^\circ.128336$, where $RA$ is the right ascension 
and $DE$ is the declination; ii) the equatorial coordinates of the galactic center: $RA = 266^\circ.405100$
and $DE = -28^\circ.936175$, evaluated at the Epoch J2000.0. In the galactic coordinates, those versors can
be written as:
\begin{eqnarray}
\hat{e}_1^{ecs} &=& (-0.05487, 0.49411, -0.86767) \nonumber \\
\hat{e}_2^{ecs} &=& (-0.87344, -0.44483, -0.19808) \\
\hat{e}_3^{ecs} &=& (-0.48384, 0.74698, 0.45599). \nonumber
\end{eqnarray}
We define $\vec{v}_{s} = \vec{v}_{LSR} + \vec{v}_{\odot}$. In this section, when a numerical calculation is employed, we
assume $v_0 = 220$ km/s; hence, $v_{s}=232.28$ km/s, and in the equatorial coordinate system: 
$\theta_{ecs} = 42^\circ.18$ (the ``zenith distance'') and $\varphi_{ecs} = - 46^\circ.14$ (azimuth angle 
from $\hat{e}_1^{ecs}$). For simplicity in the following of this section in order to offer an estimate of the {\it Earth Shadow Effect} 
diurnal behaviour we consider $\vec{v}_{rev}(t)$ equal to its annual mean value, i.e. zero. To introduce the Earth motion around its axis, 
firstly we consider the simplest case of a laboratory at North Pole and we define the horizontal coordinate 
system with $z'$ axis directed as shown in Fig.\ref{fig:fig1}, $left$ and $x'$ axis directed towards a given longitude 
$\lambda_0$. In this coordinate system ($N$ label) the velocity of the Earth can be written as: 
$v_{lab}^N = v_{s}$, $\theta_N = \theta_{ecs}$, 
and $\varphi_N = \varphi_{ecs} - \omega_{rot} (t + t_0) = -(\varphi_0 + \omega_{rot} (t+t_0))$,
where $\varphi_0 = -\varphi_{ecs}$, $\omega_{rot} = 2 \pi / T_d$ with $T_d = 1$ sidereal day, $t$ sidereal time
referred to Greenwich, and $t_0 = 24\lambda_0/2\pi$ sidereal hours.

\begin{figure}[!b]
\begin{center}
\vspace{-0.5cm}
\includegraphics[width=0.7\textwidth]{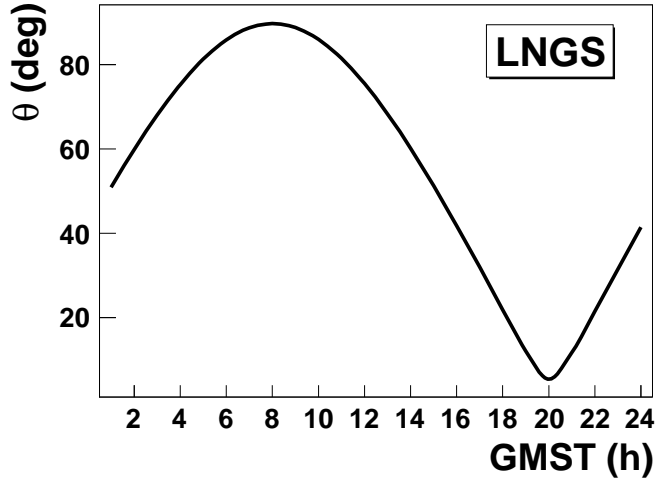}
\end{center}
\vspace{-0.7cm}
\caption{The angle $\theta$ as a function of the sidereal time in the case of the LNGS (latitude 
$\lambda_0 = 13^\circ 34'$ E, longitude $\phi_0 = 42^{\circ} 27' $ N). The Earth shielding is maximum 
about at 8:00 h and minimum around 20:00 h; see text.}
\label{fig:fig2}
\end{figure}

The general case of $\vec{v}_{lab}$ in a laboratory at latitude $\phi_0$ can be derived by rotating 
counterclockwise $\vec{v}_{lab}^N$ around $\hat{y}'$ of an angle $\alpha = \pi / 2 - \phi_0$:
\begin{eqnarray}
\vec{v}^{\lambda_0 , \phi_0}_{lab} & = & R(\alpha ) \vec{v}_{lab}^N = \\ \nonumber
& = &  \left( 
\begin{array}{ccc}
\cos \alpha & 0 & - \sin \alpha  \\
    0       & 1 &      0    \\
\sin \alpha & 0 & \cos \alpha \\
\end{array}
\right)  \left( 
\begin{array}{c}
v_{s} \sin \theta_{ecs} \cos (\varphi_0 + \omega_{rot} (t+t_0)) \\
-v_{s} \sin\theta_{ecs} \sin(\varphi_0 + \omega_{rot} (t+t_0)) \\
v_{s} \cos\theta_{ecs}
\end{array}
\right).
\end{eqnarray}
Thus, $\theta (t)$ for a laboratory position, identified by the longitude $\lambda_0$ and latitude 
$\phi_0$, can be derived from $\cos \theta (t) = \hat{v}_{lab}^{\lambda_0,\phi_0}\cdot \hat{z}' $ 
(see Fig. \ref{fig:fig1}, {\it right}) obtaining:
\begin{eqnarray}
\cos \theta (t) & = & \sin \theta_{ecs} \cos \phi_0 \cos (\omega_{rot} (t+t_0) + \varphi_0) + 
\cos \theta_{ecs} \sin \phi_0   \\
& = & \cos \psi \cos \phi_0 \cos (\omega_{rot} (t+t_0) + \varphi_0) +
\sin \psi \sin \phi_0, \nonumber
\end{eqnarray}
where $\psi = \pi/2 - \theta_{ecs} = 47^{\circ}.82$ is the altitude. In case of the LNGS, the longitude is 
$\lambda_0 = 13^\circ 34'$ E and the latitude is $\phi_0 = 42^{\circ} 27'$ N, thus $t_0 = 0.904444$ sidereal 
hours. The $\theta (t)$ behaviour at LNGS is shown in Fig \ref{fig:fig2}. Note that, before detection, the DM 
particles at LNGS preferentially cross an Earth maximum thickness at about 08:00 h and an Earth minimum 
thickness at about 20:00 h (GMST in both cases). 

\section{Deformation of the DM velocity distribution due to the {\it Earth Shadow Effect}}

To study the experimental data in terms of possible {\it Earth Shadow Effect}, a Montecarlo code has been 
developed to simulate the propagation of the DM candidates elastically scattering
off Earth nuclei in their travel in the Earth towards the 
underground experimental site. For such a purpose useful information has been gathered about the Earth 
composition and density. The Montecarlo code numerically estimates the velocity distribution 
-- in the laboratory coordinate system -- of the impinging DM particles after having crossed the Earth; such velocity 
distribution depends on the mass of the DM candidate, on its cross-section on nucleons, on the initial 
unperturbed velocity distribution, on the sidereal time, and on the latitude and longitude of the laboratory: 
$f_{lab} (v, t | m_{DM}, \sigma_n)$. Then, this velocity distribution has been used to evaluate -- in an assumed 
framework -- the expected counting rate as a function of the sidereal time in order to be compared with the 
experimental data.

In this section details are given about the assumptions adopted in the simulation, as in particular: the Earth 
model, the {\it mean free path} and path reconstruction of such DM candidate, the adopted interaction model (nuclear form 
factor, scaling law, etc.) and the $f_{lab} (v, t | m_{DM}, \sigma_n)$ estimation.

\subsubsection*{The Earth model}
An Earth model has been assumed in order to estimate the signal variation due to the {\it Earth Shadow Effect}; 
in particular, the matter density and composition of the 
Earth have to be considered. The simulation adopts a simplified Earth model starting by the 
Preliminary Reference Earth Model (PREM) \cite{Dzi81}. Just three main Earth 
layers with a constant density and homogeneous distribution in each one (these values are averaged over 
the PREM density 
distribution behaviour) are considered: i) the Inner Core; ii) the External Core; iii) the Mantle.
The densities and mass percentage for each layer are given in Table \ref{tab:tab1};
\begin{table}[!ht]
\caption{Density values, $\rho_L$, and $i$-th nucleus mass percentage, $\delta_i$, adopted in the present calculations
for the layers of the considered Earth model \cite{Dzi81}.}
\begin{center}
\begin{tabular}{|l|c|c|c|c|}
\hline
  Layer ($L$)   & R$_{min}$ -- R$_{max}$      & Mass percentage  & Density ($\rho_L$) \\ 
                &         (km)                & ($\delta_i$)     &  (kg/m$^3$)        \\\hline \hline
  Inner Core    & 0 -- 1221.5                 & Fe (79 \%)       & 12839              \\
                &                             & Ni (21 \%)       &                    \\ \hline
  External Core & 1221.5 -- 3480              & Fe (86 \%)       & 10901              \\ 
                &                             & S (12 \%)        &                    \\
                &                             & Ni (2 \%)        &                    \\ \hline
  Mantle        & 3480 -- 6371                & O (44.9 \%)      & 4605               \\ 
                &                             & Si (21.6 \%)     &                    \\
                &                             & Mg (22.8 \%)     &                    \\
                &                             & Fe (5.8 \%)      &                    \\
                &                             & Ca (2.3 \%)      &                    \\
                &                             & Al (2.2 \%)      &                    \\
                &                             & Na (0.4 \%)      &                    \\ \hline \hline
\end{tabular}
\end{center}
\label{tab:tab1}
\end{table}
in this simplified model the rare isotopes (mass percentage lower than 0.1\%) have been neglected.

\subsubsection*{Interactions of the considered DM candidates and path reconstruction}

We assume that the considered DM candidates loss their energy elastically scattering off nuclei with spin-independent coupling. 
The {\it mean free path} in the $L$-th Earth's layer is:
\begin{equation}
\lambda_L = \frac{1}{\displaystyle \sum_{i=1}^N \sigma_{DM,i} n_i},
\end{equation}
where: i) $n_i = \frac{\rho_L }{m_i}\delta_i$ is the number density of the $i$-th nuclei in the Earth's layer 
composed by $N$ nuclear species; ii) $\sigma_{DM,i}$ is the cross-section on the $i$-th nucleus and 
$m_i$ is the mass of the $i$-th nucleus; iii) $\rho_L$ is the layer's density; iv) $\delta_i$ is the $i$-th 
nucleus mass percentage in the layer. We considered a coherent scattering and a scaling law\footnote{
This equation holds in the limit where the form factor can be neglected. For  high velocity, high mass 
DM candidate crossing the Inner Core it is only an approximation. For simplicity
we do not consider further this issue here, while the form factor is considered in obtaining the
energy loss (see later).}: 
\begin{equation}
\sigma_{DM,i} = \sigma_n A_i^2 \frac{\mu_i^2}{\mu_n^2},
\label{eq:scl}
\end{equation}
where: i) $A_i$ is the mass number of nucleus $i$; ii) $\sigma_n$ is the DM candidate-nucleon cross-section; 
iii) $\mu_i$ ($\mu_n$) is the DM candidate -- nucleus(nucleon) reduced mass. Thus, assuming 
$m_i \simeq A_i m_n$, one can write:
\begin{equation}
\frac{1}{\lambda_L} = \frac{\sigma_n \rho_L}{ m_n} \sum_{i=1}^N A_i^3 \delta_i
\left( \frac{m_{DM}+m_n}{m_{DM} + A_i m_n}\right)^2. 
\label{eq:lambda}
\end{equation}

For simplicity, the deflection of the DM particles crossing the Earth is neglected
(i.e. a linear trajectory is considered); 
in this assumption the Earth thickness crossed by those DM candidates, $d$, 
depends only on the impinging angle with respect to the detector, $\theta_{in}$, according to the relation:
\begin{equation}
d = 2 R_\oplus \cos \left( \theta_{in} \right), 
\end{equation}
with $R_\oplus$ Earth radius. This $d$ value is the sum of the distances passed through by the DM candidate in each layer: $d=\sum_Ld_L$. 
Defining the maximum radii of the Earth layers $R_{ic}$, $R_{ec}$ and $R_m = R_\oplus$ for Inner Core, External Core and 
Mantle respectively (see Table \ref{tab:tab1}), the number of layers crossed by DM particles in the 
considered schema is: 
0 for $\theta_{in} \ge 90^\circ$; 
1 for $33^\circ.11 \le \theta_{in} < 90^\circ$; 
3 for $11^\circ.05 \le \theta_{in} < 33^\circ.11$; 
5 for $0^\circ \leq \theta_{in} < 11^\circ.05$ 
(considering that $\arcsin\left(R_{ec}/R_\oplus\right)=33^\circ.11$ and  
$\arcsin\left(R_{ic}/R_\oplus\right)=11^\circ.05$).

Thus, in this scenario, the DM particles move in each Earth's layer with a {\it mean free path} $\lambda_L$, that 
mainly depends on the interaction cross section $\sigma_n$ (see eq. \ref{eq:lambda}).
The number of interactions in each layer, $n_{hit}$, has been estimated as:
\begin{itemize}
\item[-] Case 1, high interaction cross sections, $d_L \geq 50 \lambda_L$: $n_{hit}$ is relatively high and
follows a gaussian distribution with mean value and variance equal to $d_L / \lambda_L$;
\item[-] Case 2, small interaction cross sections, $d_L < 50 \lambda_L$: a step-by-step approach has been 
adopted in the simulation. The path between two consecutive interactions, $x_k$, follows 
the distribution $\lambda_L^{-1} e^{-(x_k/\lambda_L)}$; it can be used to propagate the particle within the 
layer as long as $\sum_k^{n_{hit}} x_k \leq d_L$.
\end{itemize}

In the considered scenario the DM candidate particles interact via SI elastic scattering on nuclei; thus, 
their energy-loss for each interaction is given by the induced nuclear recoil energy:
\begin{equation}
E_R = E_{in} r \left( \frac{1-\cos \theta^*}{2} \right),
\label{eq:er}
\end{equation}
where $E_{in} = m_{DM} v^2 / 2$ and $v$ are the DM energy and the velocity before the interaction, 
$\theta^*$ is the angle of diffusion 
in the center of mass and $r$ is a kinematic factor:
\begin{equation}
r=\frac{4 m_{DM} m_i}{(m_{DM}+m_i)^2}.
\end{equation}
The interaction cross-section is given by \cite{allDM4}:
\begin{equation}
\frac{d\sigma_{DM,i}}{dE_R} (v,E_R) = \frac{d\sigma_{DM,i}}{dE_R} (v,0)F^2_i(E_R),
\end{equation}
with $F^2_i (E_R)$ nuclear form factor. In addition, for a given velocity $v$:
\begin{equation}
\frac{d\sigma_{DM,i}}{dE_R} (v,0) = \frac{d\sigma_{DM,i}}{d\Omega}\times \frac{d\Omega}{dE_R},
\end{equation}
with (also see eq. \ref{eq:er}):
\begin{eqnarray}
\frac{d\sigma_{DM,i}}{d\Omega} = \frac{\sigma_{DM,i}}{4\pi},  \; \; \; \frac{dE_R}{d\Omega} = \frac{rE_{in}}{4\pi}.
\end{eqnarray}
Hence, assuming the cross section scaling law given in eq. \ref{eq:scl}, the interaction cross-section can be 
written as:
\begin{equation}
\frac{d\sigma_{DM,i}}{dE_R} (v,E_R) = \frac{\sigma_{DM,i}}{r E_{in}}F^2_i(E_R) = \sigma_n A_i^2 \frac{m_i}{2 \mu_n^2 v^2}F^2_i(E_R).
\label{eq:dsig}
\end{equation}
Thus, after one interaction, such a DM particle loses an energy in the range 
$0 \leq E_R \leq r E_{in}$ with distribution given in eq. \ref{eq:dsig}. In the following we 
assume for the nuclear form factor the Helm formula\footnote{It is important to remark that
the spin independent form factor depends on the target nucleus and there is not an universal formulation for it.
Many profiles are available in literature and whatever profile needs some parameters whose value are also affected 
by some uncertainties. The form factor profiles can differ -- in some intervals of the transferred momentum -- 
by orders of magnitude and the chosen profile strongly affects whatever model dependent results \cite{RNC}.} 
\cite{Helm1,bot94}:
\begin{equation}
F_i(qr_0) = \frac{3\left[\sin (qr_0) + qr_0 \cos (qr_0) \right]}{(qr_0)^3}e^{-\frac{1}{2} q^2s^2},
\end{equation}
where $r_0 = \sqrt{r_i^2 - 5 s^2}$, $r_i = 1.2 A_i^{1/3}$ fm, $s\simeq1$ fm and $q^2 = 2m_i E_R$.

\begin{figure}[!ht]
\begin{center}
\includegraphics[width=1.0\textwidth]{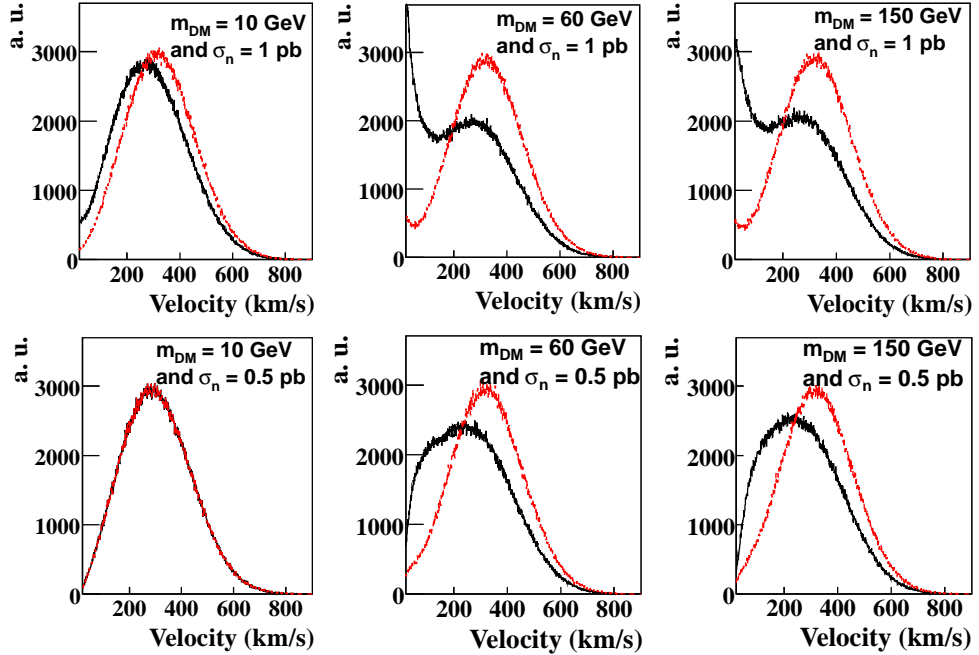}
\vspace{-1.2cm}
\end{center}
\caption{Some examples of simulated $f_{lab} (v, t | m_{DM}, \sigma_n)$ in arbitrary units (a.u.) 
for some $m_{DM}$ and $\sigma_n$ values when
taking into account the {\it Earth Shadow Effect}. In these plots: i) the considered
Galactic Halo model is an isothermal sphere with $v_0 = 220$ km/s; ii) the velocity distribution 
at GMST hour 8:00 (continuous black line) and 20:00 (dashed -- red on-line -- line) -- corresponding to the maximum and the minimum of 
the {\it Earth Shadow Effect} in case of a target-detector placed at LNGS.}
\label{fig:fig3}
\end{figure}

In the simulation, once $n_{hit}$ has been evaluated,  
the energy loss of the considered DM particle, $E_R$, is estimated for each interaction and the output velocity is:
\begin{equation}
v_f = \sqrt{ v^2 - \frac{2E_R}{m_{DM}}}.
\end{equation}
Therefore, the net effect is a modification of the velocity distribution, $f_{lab} (v, t | m_{DM}, \sigma_n)$;
Fig. \ref{fig:fig3} shows some examples of the obtained velocity distribution for a detector located at LNGS. 

\section{The expected interaction rate}

In the SI coupling scenario, considered here, the DM candidates scatter off the nuclei 
in the detector. Their expected interaction rate as a function of the nuclear recoil energy, $E_R$,  
for a mono-atomic ($i$ nucleus) detector is:
\begin{equation}
\frac{dN_i}{dE_R} = N_T \int_{v_{min}(E_R)}^\infty \frac{\rho v}{m_{DM}} f_{lab}(v) \frac{d\sigma_{DM,i}}{dE_R} (v,0) F^2_i(E_R) dv,
\end{equation}
where $N_T$ is the number density of the target nuclei in the detector and $\rho = \xi \rho_0$
is the DM particles density in the galactic halo ($\xi$ is the relative abundance and $\rho_0$ the overall DM density
in the galactic halo). The integral is calculated over all the possible DM particle velocities in the 
laboratory frame considering the distribution $f_{lab}(v) = f_{lab} (v, t | m_{DM}, \sigma_n)$.
The minimal velocity providing $E_R$ recoil energy is $v_{min}(E_R) = \sqrt{\frac{E_Rm_i}{2\mu_i^2}}$.
The galactic escape velocity is included in the $f(v)$ definition.
Using eq. \ref{eq:dsig}, the expected rate can be rewritten as:
\begin{eqnarray}
\frac{dN_i}{dE_R} &=& \xi \sigma_n \frac{N_T  \rho_0}{m_{DM}} A_i^2 \frac{m_i}{2\mu_n^2} F^2_i(E_R) 
\int_{v_{min}(E_R)}^\infty \frac{f_{lab} (v, t | m_{DM}, \sigma_n)
}{v} dv \nonumber \\
&=&  \xi \sigma_n \frac{dN_i'}{dE_R} (E_R , t | m_{DM}, \sigma_n) .
\end{eqnarray}
Generalizing to detectors with more than one kind of target nuclei (as e.g. in the case of the NaI(Tl) considered here), 
the expected experimental rate is:
\begin{eqnarray}
\frac{dR}{dE_{det}} = \xi \sigma_n \frac{dR'}{dE_{det}} (E_{det}, t | m_{DM}, \sigma_n) ,
\end{eqnarray}
where:
\begin{eqnarray}
&&\frac{dR'}{dE_{det}}(E_{det}, t | m_{DM}, \sigma_n) = \\
&&\int G(E_{det},E') \left[ \sum_i \int K_i(E'|E_R) \frac{dN_i'}{dE_R}(E_R , t | m_{DM}, \sigma_n)
dE_R \right] dE'; \nonumber 
\end{eqnarray}
the $G(E_{det},E')$ kernel takes into account the detector's energy resolution (generally through a gaussian 
convolution) and the $K_i(E'|E_R)$ kernel takes into account the energy transformation of the nuclear recoil energy
in keV electron equivalent. For example the latter kernel can be written in the simplest case of a constant
quenching factor, $q_i$, as: $K_i(E'|E_R) = \delta(E' - q_i E_R)$. For a discussion about the quenching factors see 
Refs. \cite{bot11,review}.

The expected differential rate -- as well as $f_{lab} (v, t | m_{DM}, \sigma_n)$ --
depends on the time through three different effects: i) the time dependence of 
the Earth's orbital motion velocity, $\vec{v}_{rev} (t)$ (see eq. \ref{eq:vlab}); ii) the time dependence of 
the Earth's rotation velocity around its axis, $\vec{v}_{rot} (t)$ (see eq. \ref{eq:vlab}); iii)
the possible {\it Earth Shadow Effect} which depends on $\sigma_n$. Details about the first two effects, responsible 
of the model-independent annual and diurnal modulation of the DM signal rate, respectively, 
are reported in Ref. \cite{diurna}. Following the same approach the 
expected rate in an energy interval $\Delta E_k$ can be written as\footnote{Here only the first order terms are shown (i.e. the interference
terms are omitted).}:
\begin{eqnarray}
S_k (t) & = & \int_{\Delta E_k} \xi \sigma_n \left( \frac{dR'}{dE_{det}} \right) (E_{det}, t | m_{DM}, \sigma_n) dE_{det} =\nonumber \\
        & = & \xi \sigma_n \left[ S'_{0,k} (m_{DM}, \sigma_n) + S'_{m,k}(m_{DM},\sigma_n) \cos (\omega (t-t_0)) \right. \label{eq:sk} \\
        & + & \left. S'_{d,k}(m_{DM},\sigma_n) \cos ( \omega_{rot} (t-t_d)) + S'_{d,sh,k}(m_{DM},\sigma_n,t) \right], \nonumber
\end{eqnarray}
where: i) $S'_{0,k} (m_{DM}, \sigma_n)$ is the time independent component of the expected signal;
ii) $S'_{m,k} (m_{DM},\sigma_n)$ is the annual modulation amplitude, $\omega = 2\pi /T$ with $T=1$ yr, $t_0 \simeq$ June $2^{nd}$;
iii) $S'_{d,k} (m_{DM},\sigma_n)$ is the diurnal modulation amplitude, $\omega_{rot} = 2\pi / T_d$ with $T_d = 1$ sidereal day, $t_d$ 
for the case of a detector at the Gran Sasso longitude ranges from 13.94 h to 14.07 h depending on the $v_0$ value (see Ref. \cite{diurna});
iv) $S'_{d,sh,k} (m_{DM},\sigma_n,t)$ (whose average value over $T_d$ is null) takes into account the signal variation as a function of 
the sidereal time due to a possible {\it Earth Shadow Effect}. 

The ratio $R_{dy} = S'_{d,k}(m_{DM},\sigma_n) / S'_{m,k}(m_{DM},\sigma_n) $ is model independent and
it is $R_{dy} \simeq 0.016$ at LNGS latitude; thus, considering the DAMA/LIBRA--phase1 experimental result on the DM annual modulation, 
the expected $\xi \sigma_n S'_{d,k}(m_{DM},\sigma_n)$ is order of $10^{-4}$ counts per sidereal day per kg per keV (cpd$_{sid}$/kg/keV, hereafter)
\cite{diurna}. The reached experimental sensitivity of DAMA/LIBRA--phase1 \cite{diurna} 
is not yet enough to observe such a diurnal modulation amplitude;
in fact, in the (2--4) keV energy interval considered here, 
the experimental diurnal modulation amplitude from DAMA/LIBRA--phase1 data
is $ (2.0 \pm 2.1) \times 10^{-3}$ cpd$_{sid}$/kg/keV \cite{diurna} ($< 5.5 \times 10^{-3}$ cpd$_{sid}$/kg/keV, 90\% C.L.).
Thus, in the following we do not further approach it. 

\begin{figure}[!ht]
\begin{center}
\includegraphics[width=0.7\textwidth]{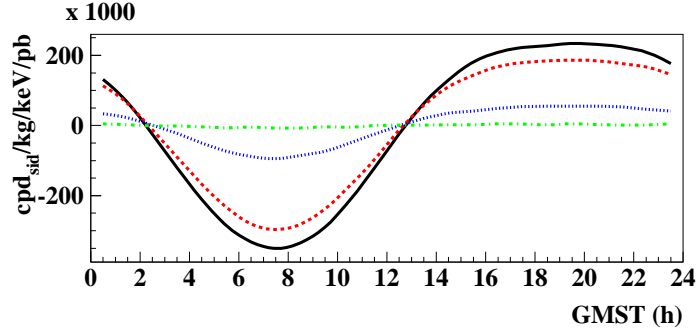}
\vspace{-0.7cm}
\end{center}
\caption{Examples of expected $S'_{d,sh,k}(m_{DM},\sigma_n,t)$ obtained following the approach described in the text. 
The Q$_I$ quenching factor and the set A of the parameters' values in presence of the {\it channeling} effect have 
been considered (see text); moreover, $v_0 = 220$ km/s. The energy interval, considered here, is (2--4) keV and the 
$m_{DM} = $ 30 GeV. The obtained $S'_{d,sh,k}(m_{DM},\sigma_n,t)$ are shown for the cases: 
i) $\sigma_n$ = 10 pb (continuous black line); 
ii) $\sigma_n$ = 1 pb (dashed -- red on-line -- line); 
iii) $\sigma_n$ = 0.1 pb (dotted -- blu on-line -- line); 
iv) $\sigma_n$ = 0.01 pb (dot-dashed -- green on-line -- line).}
\label{fig:fig4}
\end{figure}

Few examples of the $S'_{d,sh,k} (t)$ behavior for $m_{DM} = 30$ GeV
and for different values of the cross-section in the given framework are shown in 
Fig. \ref{fig:fig4}. For clarity, in the explicative case of 
$\sigma_n = 10$ pb and $\xi = 1.1 \times 10^{-8}$ so that 
$\xi \sigma_n = 1.1 \times 10^{-7}$ pb is compatible with the DAMA/LIBRA--phase1 DM annual modulation result,
the obtained amplitude of $\xi \sigma_n S'_{d,sh,k}(t)$ is of order of $3 \times 10^{-2}$ cpd$_{sid}$/kg/keV.
Such value can be studied in DAMA/LIBRA--phase1 (see later).

\section{Data analysis}

The results, obtained by analysing in the framework of the {\it Earth Shadow Effect} 
the DAMA/LIBRA--phase1 (total exposure 1.04 ton$\times$yr) data,
essentially depend on the most sensitive (2--4) keV interval; thus, this is the energy region considered
here.

In the present analysis, as in Refs. \cite{bot11,ldm}, three possibilities for the Na and I quenching factors 
have been considered: Q$_I$) the quenching factors of Na and I ``constants''
with respect to the recoil energy $E_{R}$: $q_{Na}\simeq 0.3$ and $q_{I}\simeq 0.09$ as measured by 
DAMA with neutron source integrated over the $6.5-97\, \rm keV$ and the $22-330\, \rm keV$ 
recoil energy range, respectively \cite{que}; Q$_{II}$) the quenching factors evaluated as in Ref. \cite{Tretyak}
varying as a function of $E_R$; Q$_{III}$) the quenching factors with the same behaviour of Ref. \cite{Tretyak},
but normalized in order to have their mean values  
consistent with Q$_{I}$ in the energy range considered there.
 
Another important effect is the {\it channeling} of low energy ions along axes and planes of the NaI(Tl) DAMA crystals.
This effect can lead to an important deviation, in addition to the other uncertainties discussed above. In fact, the 
{\it channeling} effect in crystals implies that a fraction of nuclear recoils are channeled and experience
much larger quenching factors than those derived from neutron calibration (see \cite{bot11,chan} for a discussion
of these aspects). 
Since the {\it channeling} effect cannot be generally pointed out with neutron measurements as
already discussed in details in Ref. \cite{chan}, only modeling has been produced up to now. In particular,
the modeling of the {\it channeling} effect described by DAMA in Ref. \cite{chan} is able to reproduce the recoil spectrum
measured at neutron beam by some other groups (see Ref. \cite{chan} for details). 
For completeness, we mention an alternative {\it channeling} model, as that of Ref. \cite{Mat08}, where larger probabilities of the
planar channeling are expected. Moreover, we mention the analytic calculation claiming that the {\it channeling} effect holds
for recoils coming from outside a crystal and not from recoils produced inside it, due to the blocking effect \cite{gelmini}.
Nevertheless, although some amount of blocking effect could be present, the precise description of the 
crystal lattice with dopant and trace contaminants is quite difficult and analytical calculations require 
some simplifications which can affect the result.
Because of the difficulties of experimental measurements and of theoretical estimate of this {\it channeling} effect,
in the following it  will be either included or not in order to give idea on the related uncertainty.

Thus, the data analysis has been repeated in some discrete cases 
which allow us to account for the uncertainties on the quenching factors and on the 
parameters used in the nuclear form factors. The first case (set A) is obtained  
considering the mean values of the parameters of the used nuclear form factors (see above and Ref. \cite{RNC}) 
and of the quenching factors. The set B adopts the same procedure as in Refs. \cite{allDM6,allDM7},
by varying (i) the mean values of the measured $^{23}$Na and $^{127}$I quenching factors
up to +2 times the errors; (ii) the nuclear radius, $r_i$, and the nuclear surface thickness parameter, $s$, 
in the form factor from their central values down to -20\%.
In the last case (set C) the Iodine nucleus parameters are fixed at the values of case B, 
while for the Sodium nucleus one considers: (i) $^{23}$Na quenching factor at the lowest value measured 
in literature; (ii) the nuclear radius, $r_i$, and the nuclear surface thickness parameter, $s$, 
in the SI form factor from their central values up to +20\%.
Finally, three values of $v_0$ have been considered: i) the mean value: 220 km/s, and ii) two extreme cases: 
170 and 270 km/s.

Because of the large number of the needed simulations, the mass of the DM candidate and of the
cross section on nucleon have been discretized as in the following: six $m_{DM}$ (5 GeV, 10 GeV, 30 GeV, 60 GeV, 100 GeV and 150 GeV)
and eight $\sigma_n$ (10 pb, 5 pb, 1 pb, 0.5 pb, 0.1 pb, 0.05 pb, 0.01 pb and 0.005 pb).

\begin{figure}[!ht]
\begin{center}
\vspace{-0.5cm}
\includegraphics[width=0.7\textwidth]{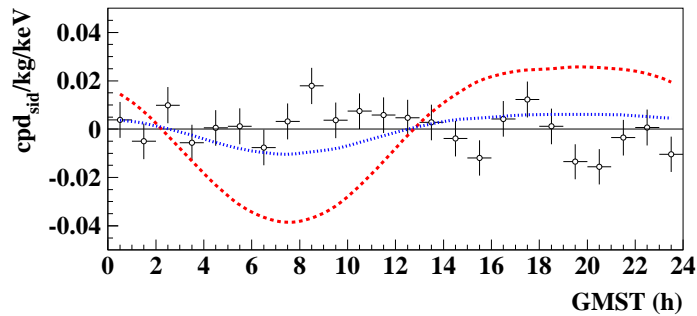}
\vspace{-0.7cm}
\end{center}
\caption{Experimental model-independent diurnal residual rate of the {\it single-hit} scintillation 
events measured by DAMA/LIBRA--phase1 in the (2--4) keV energy interval (crosses) as function of the sidereal time
\cite{diurna}, superimposed to two examples of expectations obtained by MC simulation.
The template curves are obtained in the same scenario as in Fig. \ref{fig:fig4}, considering 
$\xi \sigma_n = 1.1 \times 10^{-7}$ pb (compatible with the DAMA/LIBRA--phase1 
DM annual modulation result): i) the dashed (red on-line) line is obtained for $\sigma_n = 10$ pb 
(and, thus, $\xi=1.1 \times 10^{-8}$); ii) the 
dotted (blu on-line) line is obtained for $\sigma_n = 0.1$ pb 
(and, thus, $\xi=1.1 \times 10^{-6}$).
The latter is compatible with the absence of diurnal rate variation in DAMA/LIBRA--phase1,
while the former is not.}
\label{fig:fig5}
\end{figure}

The expectations are compared with the experimental model-independent diurnal
residual rate of the {\it single-hit} scintillation events, measured by DAMA/LIBRA--phase1 in the (2--4) keV energy interval,
as function of the sidereal time (see in Fig. 2 of Ref. \cite{diurna}).
Two examples of expected signals are reported in Fig. \ref{fig:fig5}.
Here, the used sidereal time bin is 1 hour (24 time bins in the sidereal day) and the experimental residuals are: 
$S_{d}^{exp}(t_i) \pm \sigma_{d,i}$. 
We compute the $\chi^{2}$ quantity:
\begin{eqnarray}
\label{chi2}
\chi^2 & = & \displaystyle \sum_{i=1}^{24} \frac{\left( S_{d}^{exp}(t_i) - \xi \sigma_n S'_{d,sh,k}(m_{DM},\sigma_n,t_i) 
\right)^2}{ \sigma_{d,i}^2} = \\ \nonumber
& = & \chi_0^2 - 2 B \xi + A \xi^2, 
\end{eqnarray}
where
\begin{eqnarray}
\chi_0^2 & = & \displaystyle \sum_{i=1}^{24} \frac{\left( S_{d}^{exp}(t_i) \right)^2}{ \sigma_{d,i}^2}
\nonumber  \\
B(m_{DM},\sigma_n) & = &  \sigma_n \displaystyle \sum_{i=1}^{24} \frac{S_{d}^{exp}(t_i) \times  S'_{d,sh,k}(m_{DM},\sigma_n,t_i)}
{\sigma_{d,i}^2} \nonumber \\
A(m_{DM},\sigma_n) & = &  \sigma_n ^2 \displaystyle \sum_{i=1}^{24} \frac{ \left( S'_{d,sh,k}(m_{DM},\sigma_n,t_i) \right)^2}
{\sigma_{d,i}^2}  \nonumber .
\end{eqnarray}
The $S'_{d,sh,k}(m_{DM},\sigma_n,t_i)$ have been evaluated for each set of parameters described above.

The $\chi^2$ of eq. \ref{chi2} is function of only one parameter, $\xi$.
Since the data do not show the presence of significant diurnal variation in the counting rate as already 
described in Ref. \cite{diurna}, only upper limits for $\xi$ are allowed, once given $m_{DM}$ and $\sigma_n$.
We can define:
$$\Delta \chi^{2}\{\xi\}=\chi^{2}\{\xi\}-\chi^{2}_0.$$
The $\Delta \chi^{2}$ is a $\chi^{2}$ with one degree of freedom and is used to determine the upper limit
of $\xi$ parameter at $2\sigma$ C.L..

\begin{figure}[!ht]
\begin{center}
\vspace{-0.6cm}
\includegraphics[width=0.4\textwidth]{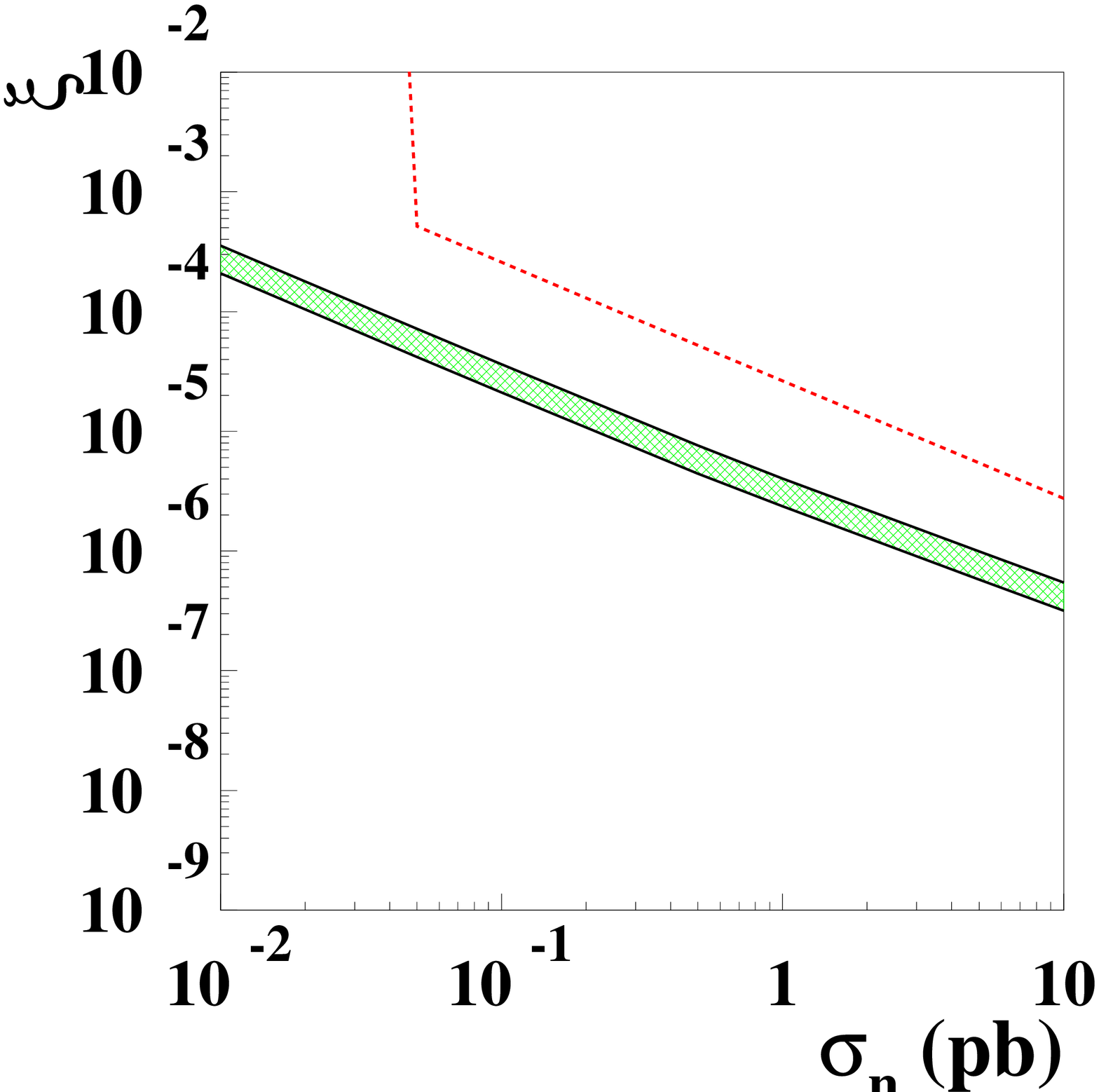}
\includegraphics[width=0.4\textwidth]{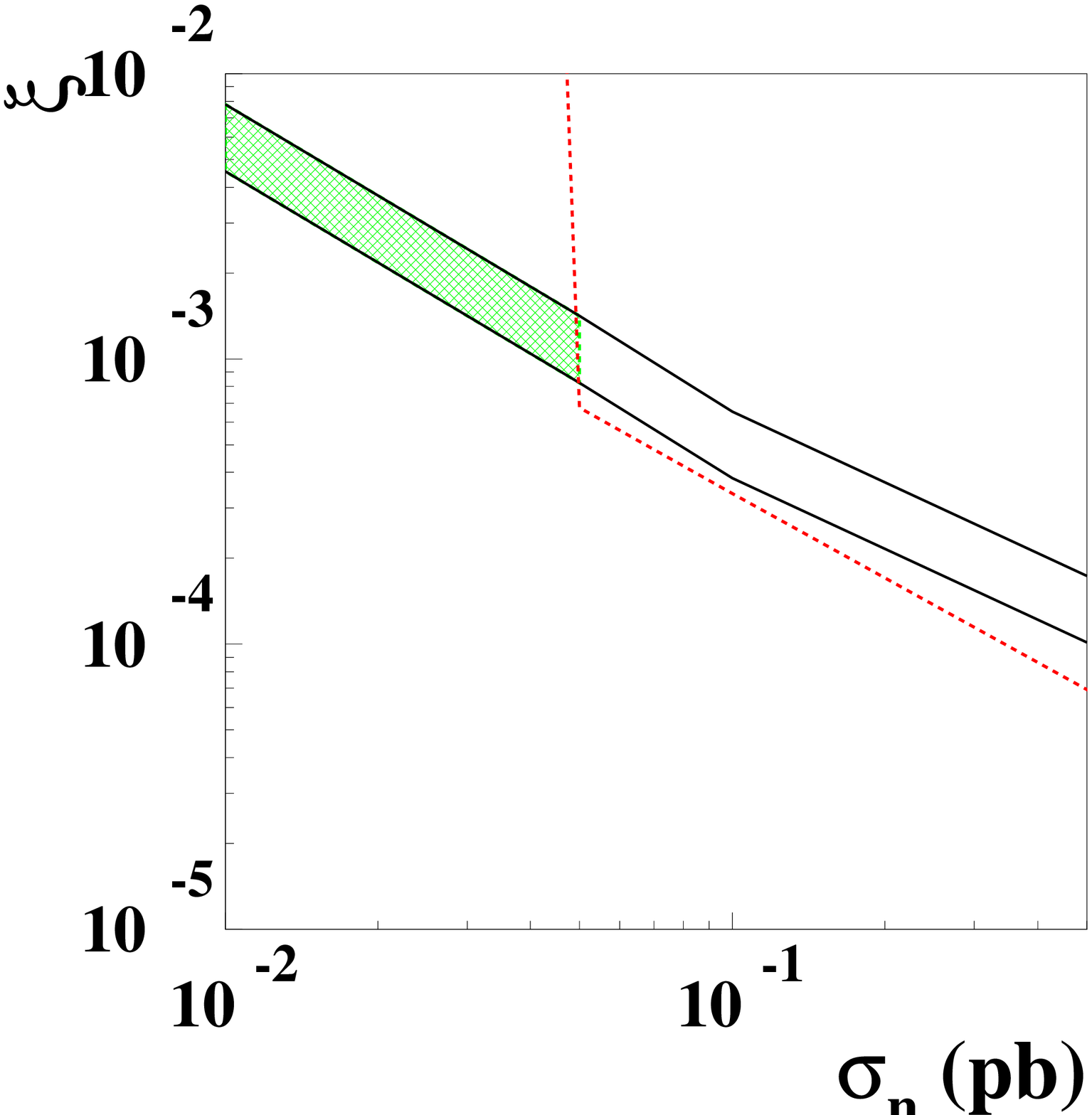}
\vspace{-0.6cm}
\end{center}
\caption{Examples of comparisons at $2\sigma$ C.L. between allowed regions from the DM annual modulation results
of DAMA/LIBRA--phase1 (black continuous lines) and exclusion limits from the {\it Earth Shadow Effect} (dotted 
-- red on-line -- line). 
In both cases the quenching factors Q$_I$, including the {\it channeling} effect, and the set B of the parameters have 
been considered
(see text) for $v_0 = 220$ km/s and for two DM masses.
({\it Left}):  $m_{DM} = 10$ GeV, the upper limits on $\xi$ do not constrain the results of annual modulation.
({\it Right}): $m_{DM} = 60$ GeV, the upper limits on $\xi$ do exclude the band with $\sigma_n > 0.05$ pb and
$\xi > 10^{-3}$ for the considered model framework.
The combined allowed regions are reported as shaded -- green on-line -- area.}
\label{fig:fig6}
\end{figure}

Two examples to describe the followed procedure are reported in Fig. \ref{fig:fig6}, where 
the excluded regions (above the dotted lines) 
in the $\xi$ vs $\sigma_n$ plane for the cases of 
$m_{DM} =$ 10 and 60 GeV are shown as obtained on the basis of the {\it Earth Shadow Effect} in the given model framework.

The upper limits on $\xi$ can be compared with the positive results from the DM annual modulation signature
achieved by DAMA\footnote{We recall that DAMA/LIBRA and the former DAMA/NaI have cumulatively reached a 
model independent evidence at 9.3$\sigma$ C.L. for the presence of DM particles in the galactic halo on the basis of 
the exploited DM annual modulation signature \cite{modlibra3}.}.
In particular, DAMA/LIBRA--phase1 reports an annual modulation amplitude in the (2--4) keV energy interval: 
$S_m^{exp} = (0.0167 \pm 0.0022)$ cpd/kg/keV, corresponding to 7.6$\sigma$ C.L. \cite{modlibra3}. 

\begin{figure}[!ht]
\vspace{-0.5cm}
\begin{center}
\includegraphics[width=0.45\textwidth]{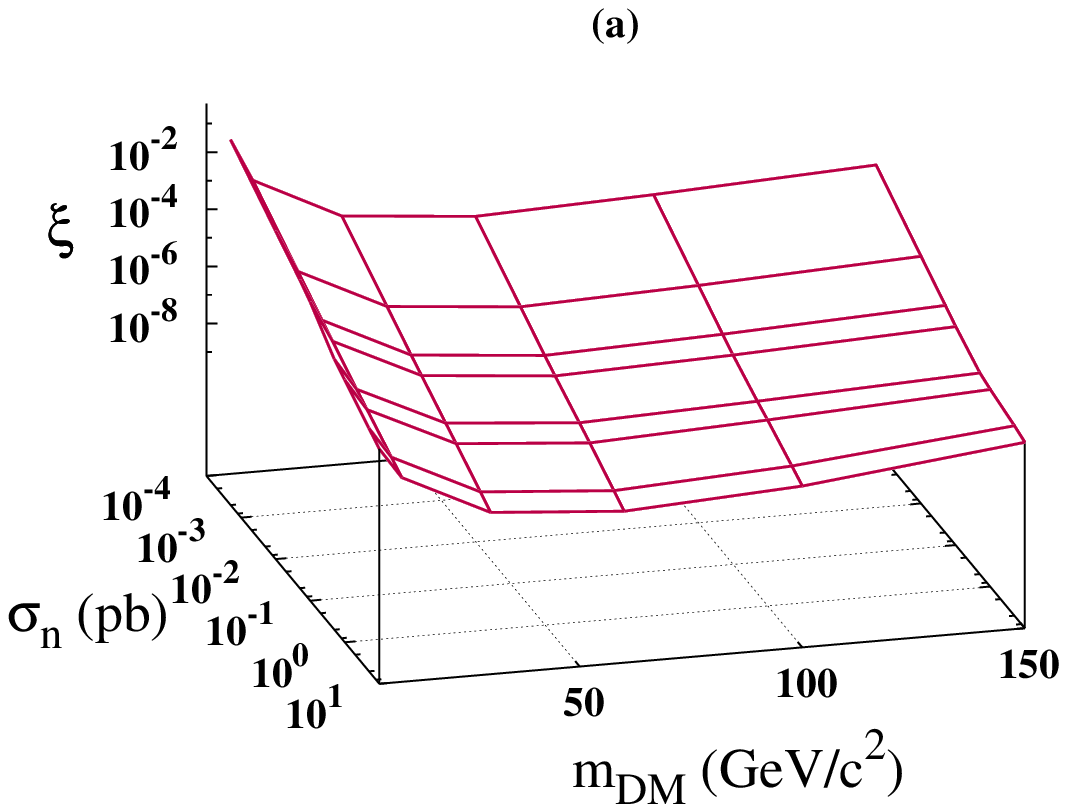}
\includegraphics[width=0.45\textwidth]{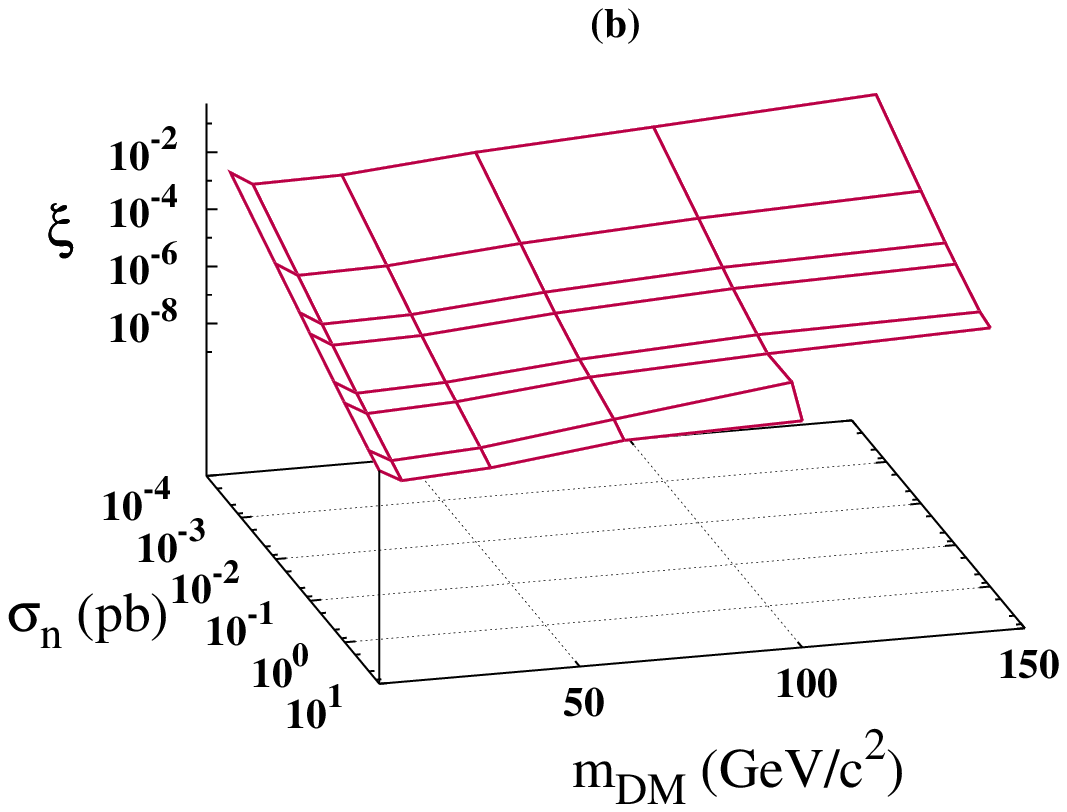}
\includegraphics[width=0.45\textwidth]{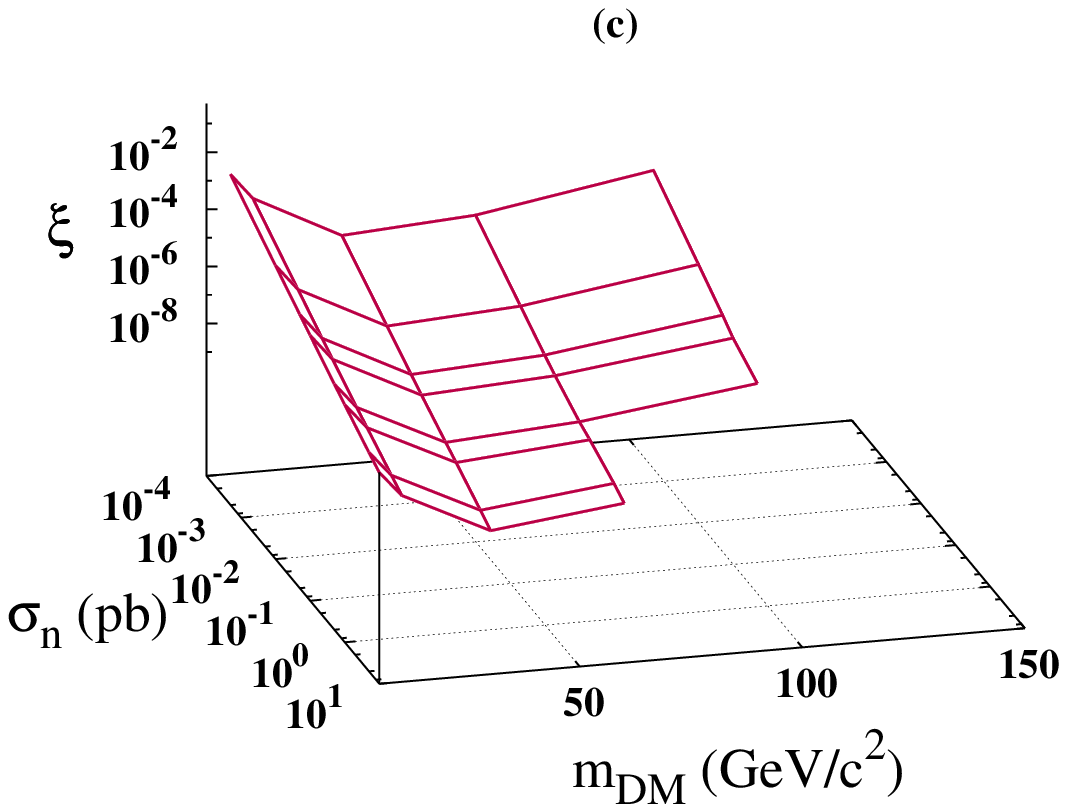}
\includegraphics[width=0.45\textwidth]{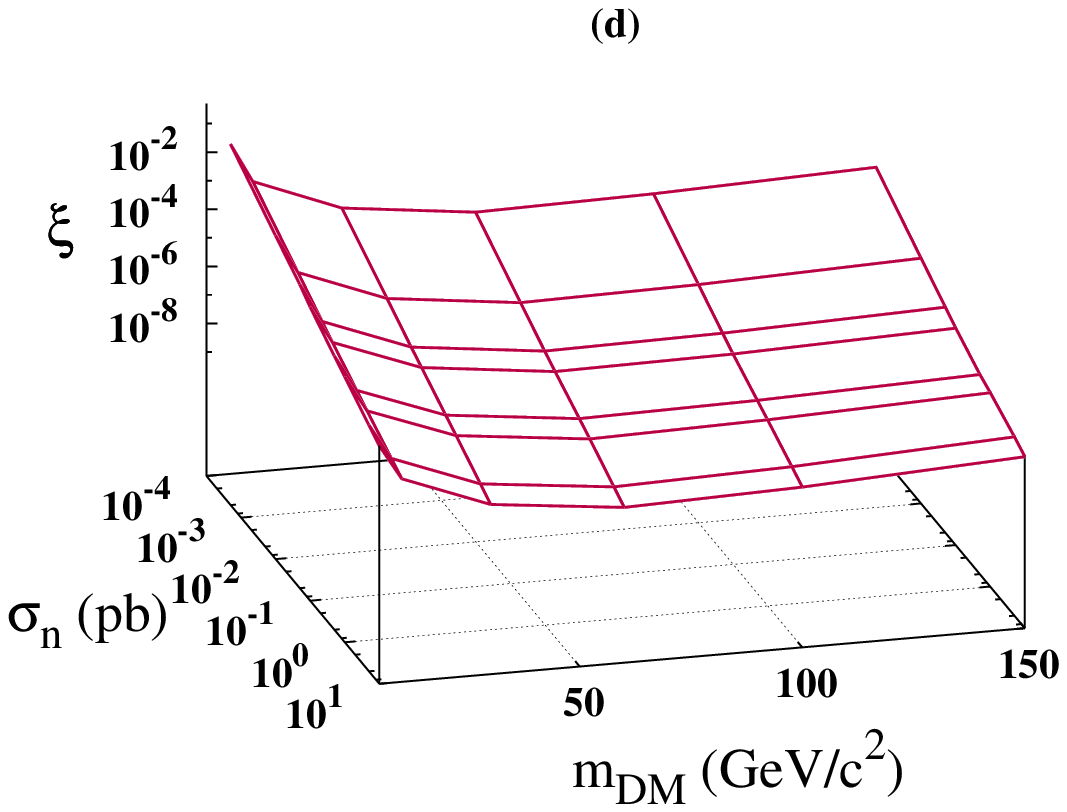}
\end{center}
%\vspace{-0.3cm}
\caption{Examples of the mean values of the allowed region of $\xi$ as function of $\sigma_n$ and $m_{DM}$,
represented as an allowed surface (see text). 
The plots have been obtained for $v_0 = 170$ km/s in the following scenarios:
a) the quenching factors Q$_I$, without {\it channeling} and marginalizing over the parameters sets A, B, C;
b) as in case a) including the {\it channeling} effect;
c) the quenching factors Q$_{II}$; 
d) the quenching factors Q$_{III}$. 
We note that the ``thickness'' of the allowed regions around the surfaces is $\leq \pm 30 \%$; 
therefore, for simplicity it is not represented in these Figures. 
Finally, we recall that other uncertainties not considered in the present paper are present and
can extend the result.}
\label{fig:fig7}
\end{figure}
\begin{figure}[!ht]
\vspace{-0.8cm}
\begin{center}
\includegraphics[width=0.45\textwidth]{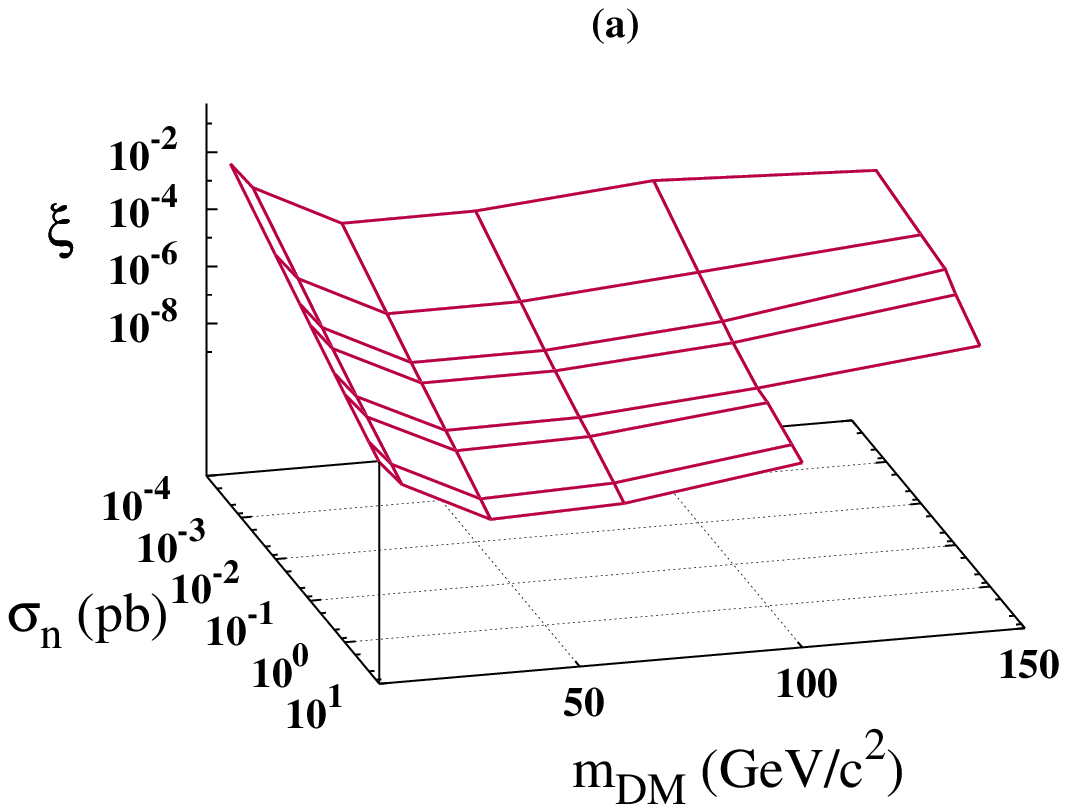}
\includegraphics[width=0.45\textwidth]{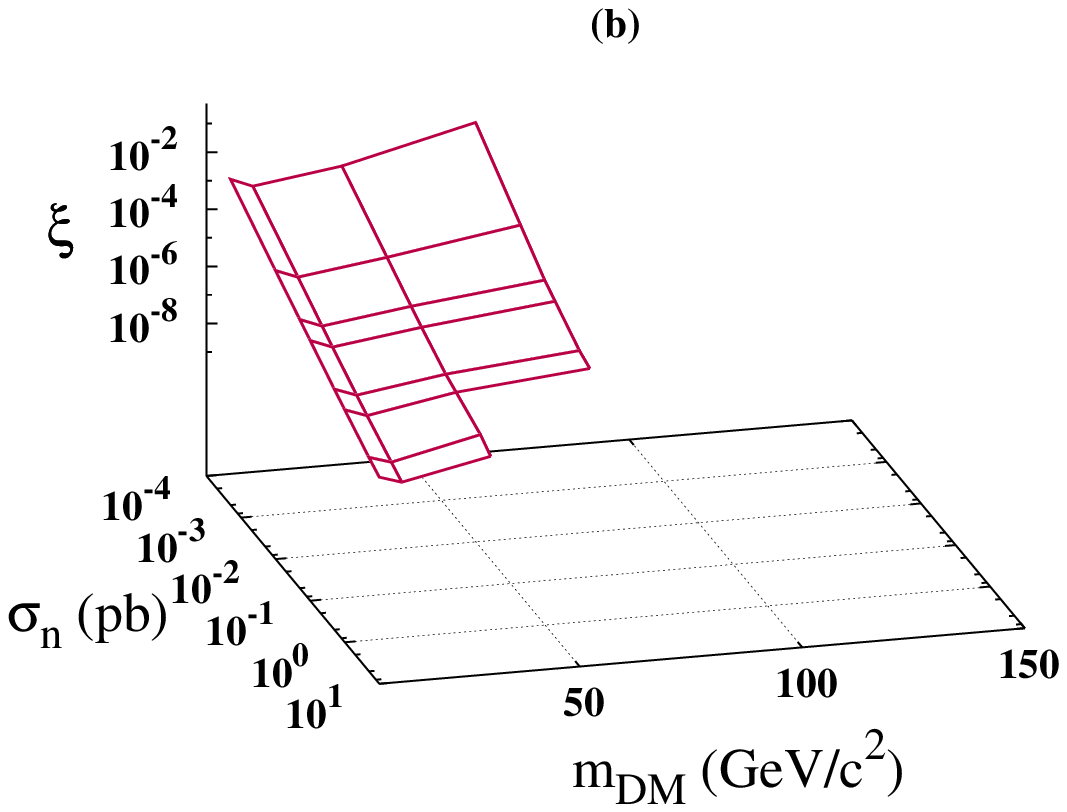}
\includegraphics[width=0.45\textwidth]{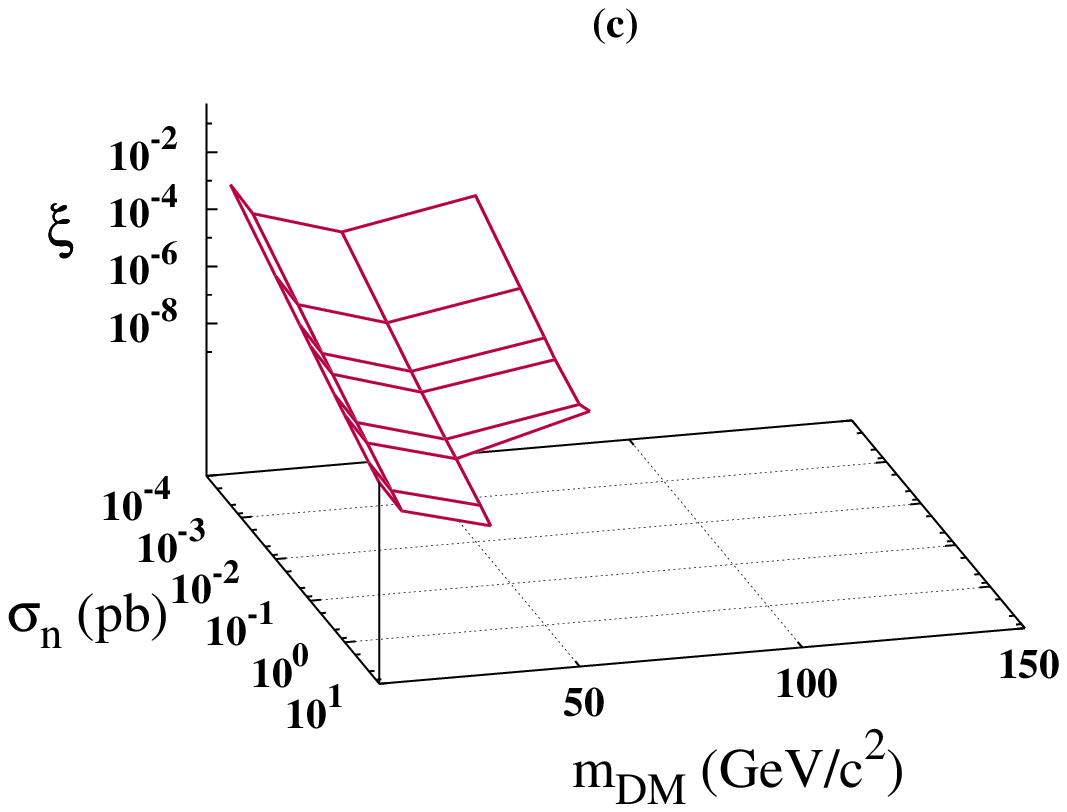}
\includegraphics[width=0.45\textwidth]{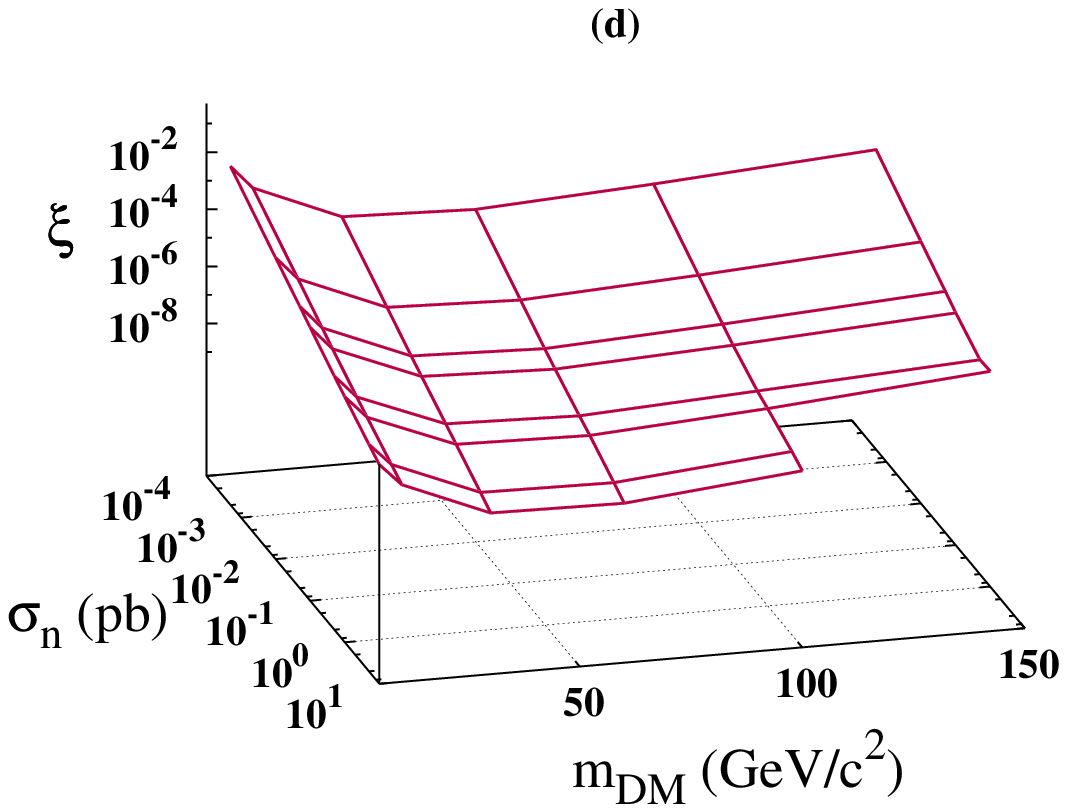}
\end{center}
\vspace{-0.6cm}
\caption{Examples of the mean values of the allowed region of $\xi$ as function of $\sigma_n$ and $m_{DM}$,
represented as an allowed surface (see text). 
The plots have been obtained for $v_0 = 220$ km/s in the same sets of parameters as in Fig. \ref{fig:fig7}.
See text.}
\vspace{-1.0cm}
\label{fig:fig8}
\end{figure}
\begin{figure}[!hb]
\vspace{-0.8cm}
\begin{center}
\includegraphics[width=0.45\textwidth]{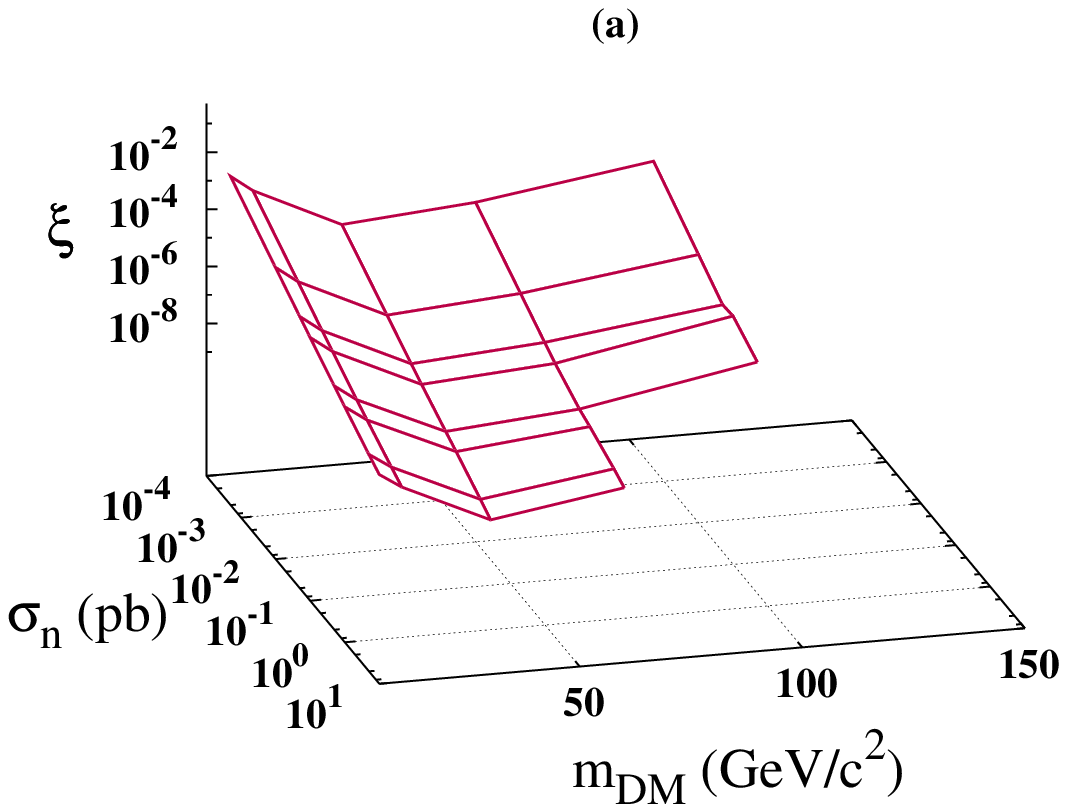}
\includegraphics[width=0.45\textwidth]{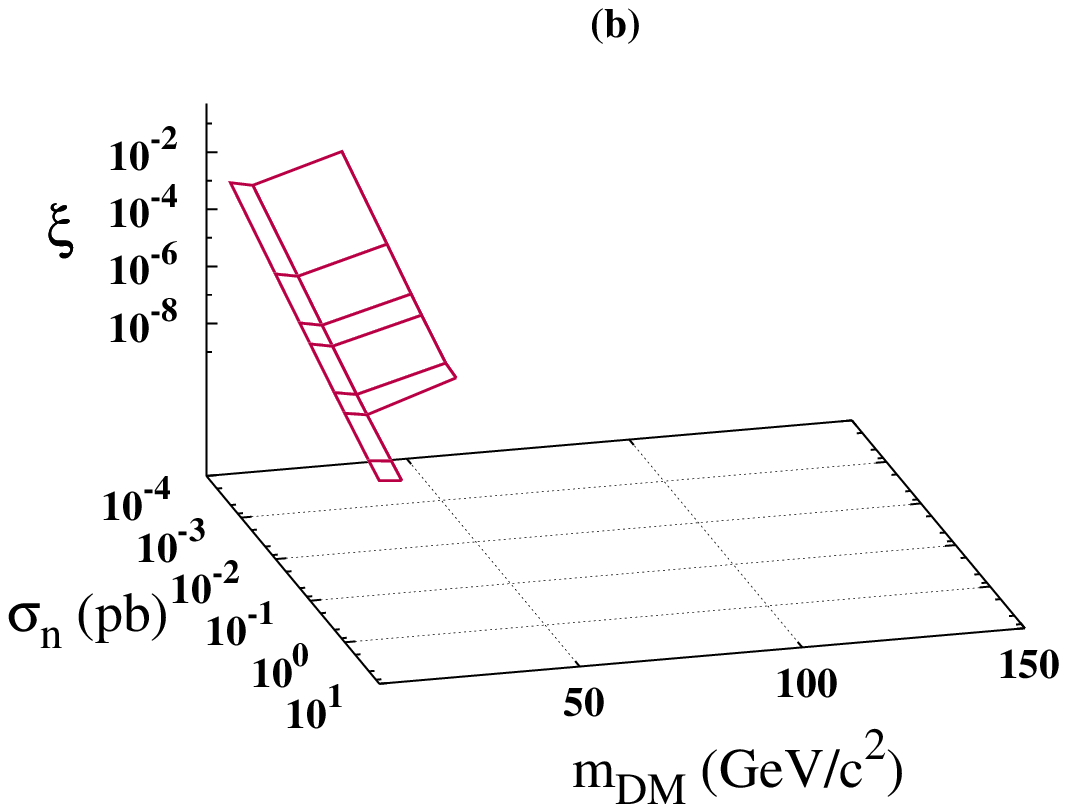}
\includegraphics[width=0.45\textwidth]{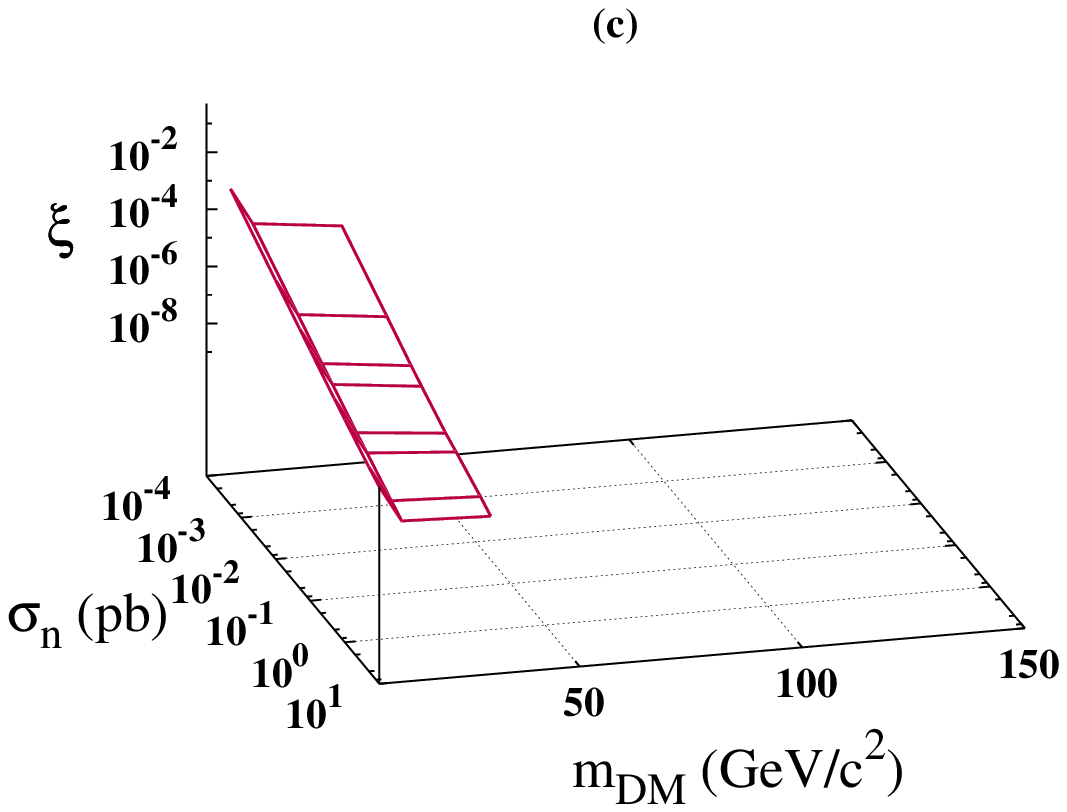}
\includegraphics[width=0.45\textwidth]{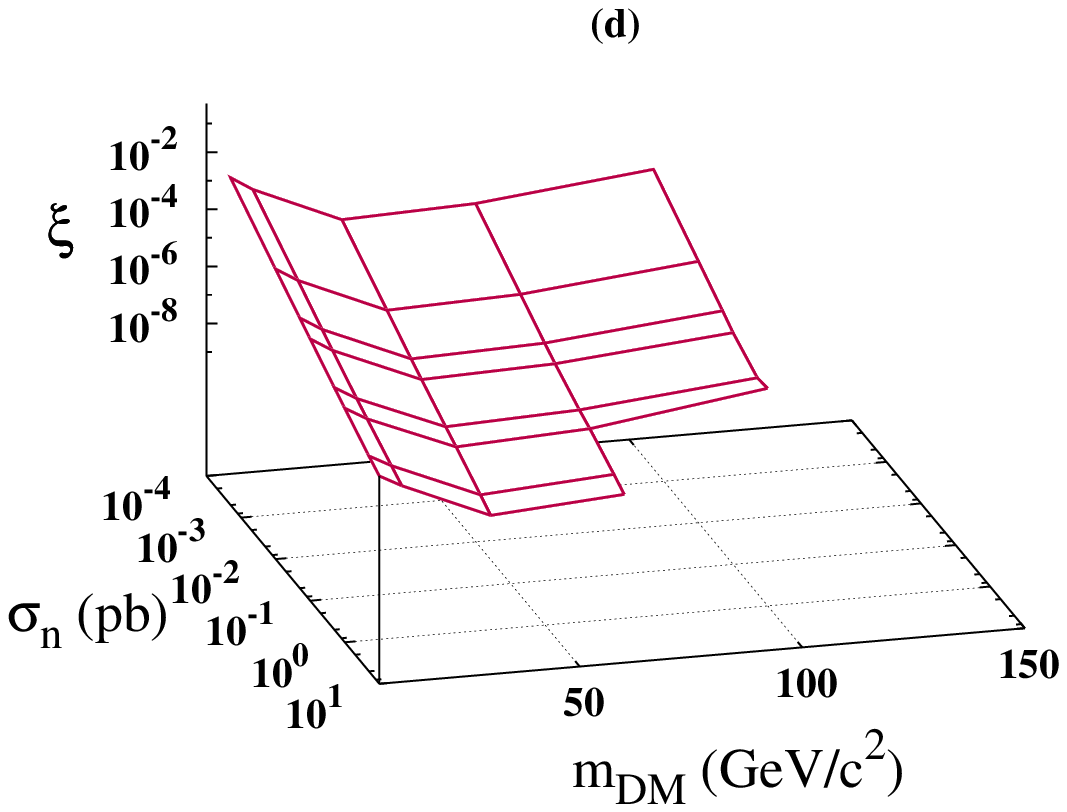}
\end{center}
\vspace{-0.6cm}
\caption{Examples of the mean values of the allowed region of $\xi$ as function of $\sigma_n$ and $m_{DM}$,
represented as an allowed surface (see text). 
The plots have been obtained for $v_0 = 270$ km/s in the same sets of parameters as in Fig. \ref{fig:fig7}.
See text.}
\label{fig:fig9}
\end{figure}

Here for each set of parameters described above, one can evaluate (see e.g. eq. \ref{eq:sk}) 
the $\xi \sigma_n$ allowed values as:
\begin{equation}
\xi \sigma_n = \frac{S_m^{exp}}{S'_{m,(2-4) keV}(m_{DM},\sigma_n) }.
\end{equation}
This corresponds, once including the experimental uncertainties on $S_m^{exp}$, to a band in the $\xi$ vs $\sigma_n$ plane
(within the continuous solid line).
In Fig. \ref{fig:fig6} such bands at $2\sigma$ C.L. are reported. One can see that for the scenario considered there
and for $m_{DM}=$ 10 GeV the upper limits on $\xi$ do not constrain the results of the DM annual modulation.
On the contrary, for $m_{DM}=$ 60 GeV the upper limits on $\xi$ do exclude the band with $\sigma_n > 0.05$ pb and
$\xi > 10^{-3}$. The shaded bands in Fig. \ref{fig:fig6} corresponds to the allowed regions in the $\xi$ vs $\sigma_n$ plane,
for the given $m_{DM}$, from the combined analyses of the DM annual modulation result and of the {\it Earth Shadow Effect}
in the considered framework.

Finally, for each considered set of parameters the three-dimensional allowed region -- calculated as described above -- in the 
parameter's space: $\xi$, $\sigma_n$, $m_{DM}$, is depicted as a surface in Fig. \ref{fig:fig7}, \ref{fig:fig8} and \ref{fig:fig9}, for
$v_0$ equal to 170, 220, 270 km/s, respectively. 
We note that the ``thickness'' of the allowed regions around the shown surfaces is $\leq \pm 30 \%$; 
therefore, for simplicity it is not represented in these Figures.

Finally, we recall that other uncertainties not considered in the present paper are present.
For example, including other possible halo models sizeable differences 
are expected in the results as shown e.g. in Refs. \cite{bot11,halo}.

\section{Conclusions}

The {\it Earth Shadow Effect} has been investigated in a given framework  
considering the model independent results on possible diurnal 
variation of the low-energy rate of the {\it single-hit} scintillation events in
the DAMA/LIBRA--phase1 data (exposure: 1.04 ton$\times$yr) reported in Ref. \cite{diurna}.
For the considered DM candidates having high interaction cross-sections and very small halo fraction
the obtained results 
constrain at 2$\sigma$ C.L., in the considered scenario, the $\xi$, $\sigma_n$ and $m_{DM}$ parameters 
(see Figs. \ref{fig:fig7}, \ref{fig:fig8} and \ref{fig:fig9}) 
when including the positive results from the DM annual modulation analysis of the DAMA/LIBRA--phase1 data 
\cite{modlibra3}.
For example, in the considered scenario for quenching factors Q$_I$ with {\it channeling} effect, 
B parameters set, $v_0 = 220$ km/s and $m_{DM} = 60$ GeV, 
the obtained upper limits on $\xi$ do exclude $\sigma_n > 0.05$ pb and $\xi > 10^{-3}$.
When also including other uncertainties as other halo models etc. the results would be extended.

\end{document}